# Effect of the Optical Pumping and Magnetic Field on the States of Phase Separation Domains in $Eu_{0.8}Ce_{0.2}Mn_2O_5$

E. I. Golovenchits[a], B. Kh. Khannanov[a], and V. A. Sanina[a],*

[a] Ioffe Institute, St. Petersburg, 194021 Russia
*e-mail: sanina@mail.ioffe.ru

The effect of optical pumping and applied magnetic field on the characteristics of ferromagnetic layers in one-dimensional superlattices is studied. At low enough temperatures, these layers correspond to phase separation domains in $RMn_2O_5$ and $R_{0.8}Ce_{0.2}Mn_2O_5$ multiferroics. The formation of such domains occurs owing to the charge ordering of $Mn^{3+}$ and $Mn^{4+}$ ions and to the finite probability for $e_g$ electrons to tunnel between these pairs of ions. The volume occupied by such superlattices is rather small, and they can be treated as isolated ferromagnetic semiconductor heterostructures, spontaneously formed in the host crystal. The sequences of ferromagnetic resonances related to the superlattice layers in $Eu_{0.8}Ce_{0.2}Mn_2O_5$ are studied. The characteristics of these resonances give information on the properties of such layers. For the first time, it is demonstrated that the optical pumping gives rise to a new metastable state of superlattices, which can be recovered by the magnetic field cycling to the state existing before the optical pumping. It is found that the superlattices recovered by the magnetic field exist up to temperatures higher than those in as-grown crystals.

The presence of equal numbers of $Mn^{3+}$ and $Mn^{4+}$ manganese ions is a characteristic feature of $RMn_2O_5$ multiferroic compounds (*Pbam* symmetry), ensuring the formation of charge ordering. $Mn^{4+}$ ions are surrounded by oxygen octahedra and located in the layers having the positions $z = 0.25c$ and $(1 - z) = 0.75c$. These ions contain three $t_{2g}$ electrons and have an empty degenerate orbital doublet corresponding to the $e_g$ state. $Mn^{3+}$ ions contain three $t_{2g}$ electrons in their $3d$ shell and one $e_g$ electron in the orbital doublet. These ions are located in the layers having the positions $z = 0.5c$, and their local environment in the form of pentagonal pyramids has no center of symmetry. $R^{3+}$ ions with the environment similar to that of $Mn^{3+}$ are located in the $z = 0$ layers [1]. Charge ordering and the finite probability of the $e_g$ electron transfer between neighboring $Mn^{3+}$–$Mn^{4+}$ ion pairs (double exchange [2, 3]) are key factors determining the multiferroic characteristics of $RMn_2O_5$ compounds in the range from low to room temperatures. Magnetic ordering with a complicated structure occurs below the Neel temperature $T_N \approx 40-45$ K, whereas the ferroelectric ordering induced by the magnetic order arises below the Curie temperature $T_C \approx 35-40$ K [4]. The polar order is mainly related to the exchange striction induced by neighboring $Mn^{3+}$ and $Mn^{4+}$ ion pairs located along the *b* axis and having alternating ferromagnetic and antiferromagnetic spin orientations [5]. The $e_g$ electron transfer between $Mn^{3+}$ and $Mn^{4+}$ ions located in adjacent layers perpendicular to the *c* axis leads to the formation of local polar regions of phase separation with a different distribution of $Mn^{3+}$ and $Mn^{4+}$ ions as compared to that in the initial host crystal. The states resulting from phase separation were studied in detail in pure $RMn_2O_5$ multiferroic manganites, as well as in the doped ones, $R_{0.8}Ce_{0.2}Mn_2O_5$ (R = Eu, Gd, Bi, Tb, and Er), having the same *Pbam* symmetry [6-13]. In these materials, the comparative studies of dielectric and magnetic properties, heat capacity, X-ray diffraction, Raman light scattering [6,7], electric polarization [8-11], and µ-*SR* [12, 13] were carried out.

Phase separation was observed in undoped $RMn_2O_5$ crystals, whereas their doping with $Ce^{4+}$ ions substituting $R^{3+}$ ions leads to a significant increase in the density of neighboring $Mn^{3+}$–$Mn^{4+}$ ion pairs in the planes perpendicular to the *c* axis. In $R_{0.8}Ce_{0.2}Mn_2O_5$ and in the $z = 0$ plane, the $R^{3+} = Ce^{4+} + e$ reaction provides additional electrons. These electrons transform $Mn^{4+}$ ions to $Mn^{3+}$ ones in the $z = 0.25c$ and $1 - z = 0.75c$ planes. As a result, the number of $Mn^{3+}$–$Mn^{4+}$ pairs and the number of phase separation domains increase. At the same time, the phase separation domains occupy a small part of the crystals both in $RMn_2O_5$ and in $R_{0.8}Ce_{0.2}Mn_2O_5$ [6, 7]. Phase separation domains are formed in $RMn_2O_5$ and in $R_{0.8}Ce_{0.2}Mn_2O_5$ in much the same way as those in $LaAMnO_3$ (A = Sr, Ca, and Ba) compounds, which also contain $Mn^{3+}$ and $Mn^{4+}$ ions [3, 14–16]. These domains are formed because of the balance between strong interactions occurring in the subsystem of Mn ions, such as the double exchange with the characteristic energy of about 0.3 eV, the Jahn−Teller interaction (0.7 eV), and the Coulomb repulsion (1 eV). That is why these domains in $RMn_2O_5$ and in $R_{0.8}Ce_{0.2}Mn_2O_5$ exist in a wide temperature range from low temperatures to those above room temperature. At $T < 60$ K, phase separation domains are isolated one-dimensional (1D) superlattices consisting of ferromagnetic layers containing $Mn^{3+}$ and $Mn^{4+}$ ions with different relative densities, as well as $e_g$ electrons changing their charge (Fig. 1). These domains manifest themselves as a series of ferromagnetic resonances (FMR) related to the individual layers of the superlattices. The characteristic features of such resonances reveal the properties of these layers and of the superlattice as a whole [17–20].

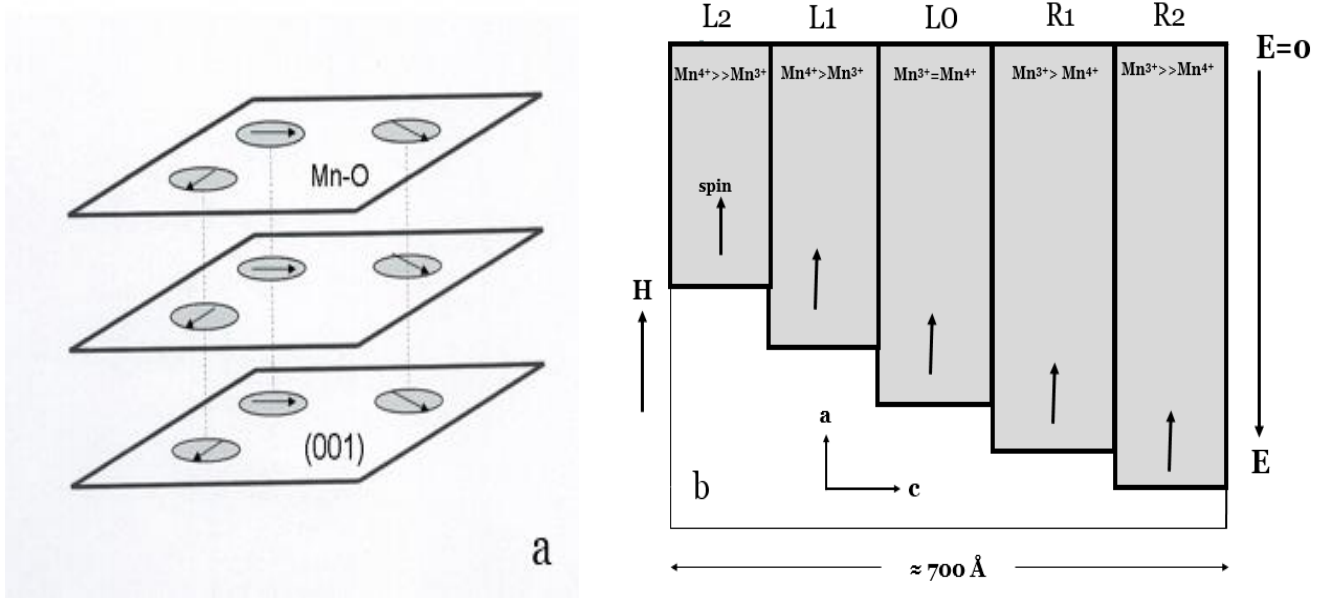

**Fig. 1.** (a) Schematic image of 1D ferromagnetic superlattices (shaded areas) located in the bulk of host crystal (white background); magnetic field $H = 0$. (b) Schematic image of one of such superlattices consisting of $L_N$ ferromagnetic layers perpendicular to the $c$ axis having different densities of $Mn^{3+}$–$Mn^{4+}$ ion pairs and of $e_g$ electrons located in potential wells of different depths (shaded areas corresponding to different energies $E$). In $Eu_{0.8}Ce_{0.2}Mn_2O_5$, the linear size of domains with superlattices is about 700 Å [6, 7].

The aim of this work is to study the effect of optical pumping, applied magnetic field, and temperature on the properties of 1D superlattices in Eu0.8Ce0.2Mn2O5 single crystals. Optical pumping can affect the electron density in the layers of superlattices and, thus, result in its redistribution in these layers. This, in turn, should change the ratio of the numbers of ions with different valences in the layers, which can be detected by changes in the characteristics of the FMR lines.

$Eu_{0.8}Ce_{0.2}Mn_2O_5$ single crystals were grown by spontaneous crystallization from the solution-melt [21, 22]. They were platelets 1−3 mm thick with an area of 3−5 mm$^2$. For FMR measurements, we used a transmission type magnetic resonance spectrometer with a low-amplitude magnetic modulation. The measurements were performed in the temperature range of 13‑300 K at frequencies of 30‑40 GHz in an applied magnetic field up to 2 T generated by an electromagnet. The cryostat with optical windows was located in the microwave channel ensuring a uniform distribution of the microwave field close to the sample. The microwave radiation (with the wave vector **k**) was directed along the *c* axis perpendicular to the platelet plane. The static magnetic field *H* was oriented along the *a* axis of the crystal and perpendicular to the direction of the microwave field *h*. The measured FMR signals were amplified by an SR530 lock-in amplifier. Naturally faceted single crystals were used. The symmetry of the crystals and their composition were determined by the X-ray phase analysis and X-ray fluorescence technique, respectively. Optical pumping was performed by a solid-state pulsed neodymium laser LTIPCH-8 with the simultaneous generation of the first (1.06 μm) and second (532 nm) harmonics. The second harmonic corresponds to the $^7F_0$–$^5D_1$ transition in $Eu^{3+}$ ions. The $^5D_1$ state is located within the electron-phonon band of Mn ions in $RMn_2O_5$, with the edge beginning at about 485 nm [23].

Earlier, the FMR studies of layers in 1D superlattices in a series of $RMn_2O_5$ (R = Eu, Er, Tb, and Gd) and $R_{1-x}Ce_xMn_2O_5$ (R = Eu and Gd, x = 0.2, 0.25) crystals demonstrated that the magnetic fields at which individual FMR lines are observed differ only slightly and are independent of the R ion type [17‑20]. In this case, $Er^{3+}$ and $Tb^{3+}$ (unlike $Gd^{3+}$) are strongly coupled to the lattice. This means that the FMR resonance fields are determined by the internal fields far exceeding the anisotropy fields characteristic of the superlattice layers; the latter fields should depend on specific R ions. Indeed, as mentioned above, the equilibrium state of 1D layers in the superlattices is determined by the balance of interactions just within the subsystem of Mn ions. In this case, the layers in the 1D superlattices can be treated as isotropic ferromagnetic layers with ferromagnetic boundaries between them, which do not put obstacles in the way of the $e_g$ electron transfer between the layers related to the double exchange (Fig. 1b) [17‑20]. A model describing the spin-wave excitations in superlattice layers was developed in these works. The model is based on the dispersion equation for spin waves in isotropic ferromagnetic films forming 2D multilayers [24]. It is also demonstrated that the FMR lines in 1D superlattices are observed in individual frequency bands separated by the ranges in which the FMR is not observed. This occurs because the 1D superlattices (similar to semiconductor superlattices [25]) have a band structure consisting of minibands separated by gaps. The frequency of 34.5 GHz, at which the measurements were performed, is located inside one of the main minibands in $Eu_{0.8}Ce_{0.2}Mn_2O_5$ (29‑36 GHz) [17].

It was reported in [17, 18] that the formation of the equilibrium state of superlattices for $EuMn_2O_5$ and $Eu_{0.8}Ce_{0.2}Mn_2O_5$, in which a set of FMR lines was observed, required the magnetic cycling (subsequently increasing and decreasing the magnetic field) of as-grown samples. The dynamically equilibrium (ground) state of the superlattices is achieved after triple magnetic cycling in the range of 0–20 kOe. Note that the ground state of the superlattices was recovered upon the cooling of the sample to helium temperatures after its heating to room temperature and its long (several weeks) keeping at this temperature. The repeated cycling of the magnetic field is not necessary.

For the 1-mm-thick $Eu_{0.8}Ce_{0.2}Mn_2O_5$ sample with an area of 5 mm$^2$, the sequence of FMR lines in the equilibrium state is shown in Fig. 2a. Generally speaking, these lines correspond to those measured earlier in [17, 18]. We can see five FMR lines originating from different layers of superlattices schematically shown in Fig. 1b. The most intense central L0 line appears in the applied magnetic field corresponding to the FMR at this frequency for an isotropic ferromagnet with the $g$ factor $g = 2$. Less intense lines, to the left of the main resonance (L1 and L2) and to the right of it (R1 and R2), are approximately symmetrically located in lower and higher magnetic fields with respect to that corresponding to the L0 line. In this case, the most intense lines (L1, L0, and R1) are doubled. At somewhat lower fields, less intense wide lines are observed (Fig. 2a).

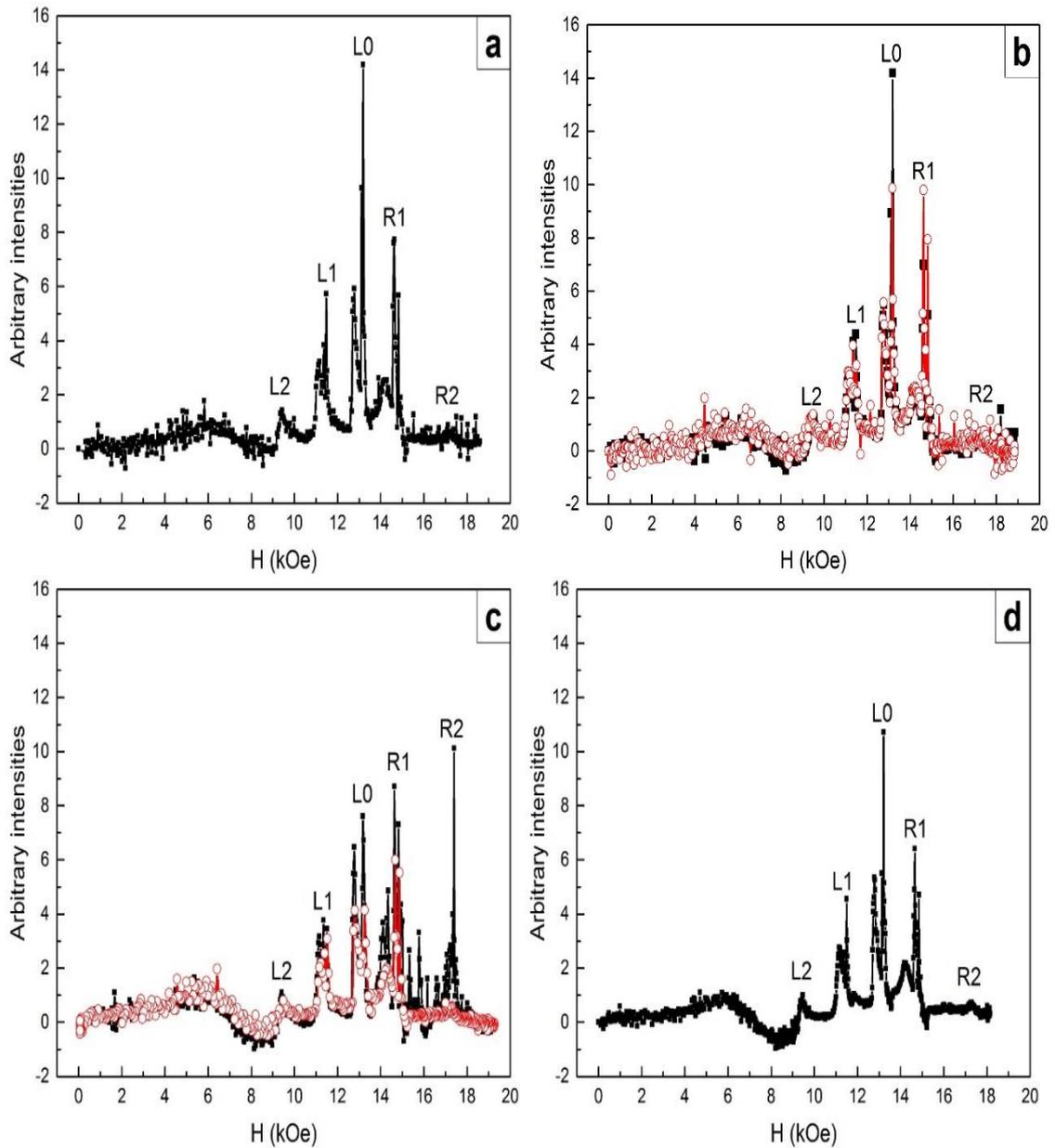

**Fig. 2.** Distribution of intensities for the observed FMR lines in $Eu_{0.8}Ce_{0.2}Mn_2O_5$ at 34.5 GHz. The sensitivity of the amplifier is 5 μV; $H \parallel a$. The magnetic field was varied at a rate of 1.2 kOe/min.

(a) As-grown sample after the third cycle of the magnetic field increase (dynamically equilibrium state of a 1D superlattice); $T = 20$ K.

(b) Dynamically equilibrium state of the superlattice (black curve) is compared to the distribution of intensities for the FMR lines after optical pumping for 1 min by 15 ns pulses with a power of 0.5 MW and a repetition rate of 10 Hz. Magnetic field $H$ decreases (red curve); $T = 22$ K.

(c) Changes in the state of the sample after 16 h of its natural heating up to 250 K and its subsequent cooling down to 17.5 K, with an increase in $H$ (black curve) and with its subsequent decrease (red curve).

(d) Recovery of the state similar to the initial dynamically equilibrium one (Fig. 2a) during the third cycle of the magnetic field growth; $T = 17.5$ K.

In [17, 18], the FMR lines were observed at temperatures below 60 K. Their intensity decreases gradually with an increase in the temperature. The resonant magnetic fields for a sequence of FMR lines exhibit a linear frequency dependence $\omega_n = \gamma_n(H + Hn^{eff})$. Here, $\omega_n$ are the circular frequencies for $n$ lines ($n$ = L2, L1, L0, R1, and R2), $\gamma_n$ are the gyromagnetic ratios, and $Hn^{eff}$ are the internal effective fields responsible for the gaps in $\omega_n(H)$. The values $Hn^{eff}$ are positive for lines L1 and L2 and increase with the number L. For lines R1 and R2, the values $Hn^{eff}$ are negative. The $g$ factors for the lines R1 and R2 are slightly smaller than 2, whereas the $g$ factors for L1 and L2 lines exceed 2 [17, 18].

Following the approach reported in [18–20], we analyze the properties of superlattice layers using the model of semiconductor heterostructures containing alternating L, 0, and R layers, in which $Mn^{4+}$ and $Mn^{3+}$ ions are treated as acceptor and donor impurities, respectively. Analyzing the characteristics of the observed FMR lines related to the layers of superlattices, we revealed the properties of these layers. We assume that the L0 layers contain equal numbers of $Mn^{4+}$ and $Mn^{3+}$ ions; i.e., these layers are fully compensated semiconductors. In such layers, the Fermi level is located in the center of the band gap, and the insulating state arises in the L0 layers [26]. The absence of free electrons in the L0 layers explains the maximum intensity of the L0 lines in the equilibrium states of superlattices (Fig. 2a). The L and R layers are partially compensated semiconductors: $Mn^{4+}$ ions play the dominant role in the L layers, whereas the Jahn–Teller $Mn^{3+}$ ions responsible for local structural distortions prevail in the R layers. The R layers exhibit the deepest potential wells with the highest electron density (Fig. 1b). According to [17–19], the R layers are characterized by negative $Hn^{eff}$ values and by $g$ factors below 2. As mentioned above, for L layers the values $Hn^{eff}$ are positive and the corresponding $g$ factors slightly exceed 2. The dynamic equilibrium resulting from the balance of competing interactions (double exchange and the Jahn–Teller effect, increasing the electron density in the layers of the superlattices, and the Coulomb repulsion of electrons), as well as the charge neutrality of the superlattice as a whole, suggests that the states of the layers should be mutually correlated and form a periodic (L−O−R) array. That is why the resonance fields of individual FMR lines turn out to be strictly fixed and almost the same in RMn$_2$O$_5$ and R$_{0.8}$Ce$_{0.2}$Mn$_2$O$_5$ compounds with different R ions [18–20].

Let us now discuss the effect of optical pumping on the state of layers in the superlattices. Pumping was carried out for 1 min at $T$ = 22 K by 15-ns pulses with a peak power of about 0.5 MW and a repetition rate of 10 Hz. In Fig. 2b, we show a sequence of FMR lines (red points on the curve) measured at the same temperature a few minutes after the end of optical pumping. These lines arise with an increase in the magnetic field at a rate of 1.2 kOe/min. These data are plotted against the background of the equilibrium state of the layers in the superlattice (black points and the curve). The sequence of FMR lines is observed at the same resonance fields as that before pumping, but the intensities of the L0 and R1 lines change. The intensities of other lines remain almost unchanged. Before pumping, the L0 line is twice as intense as the R1 line (Fig. 2a), whereas the intensities of these lines after pumping become equal (Fig. 2b). We suppose that, upon relaxation of optical excitations in the electron-phonon band related to Mn ions, electrons appearing in the layers L0 and R1 of superlattices turn out to be in excess as compared to the initial electron density formed because of doping with Ce$^{4+}$ ions. Most of these electrons are localized at Mn$^{4+}$ ions (Mn$^{4+}$ + $e$ = Mn$^{3+}$), transforming them to Jahn–Teller ions, which lead to local lattice distortions in the corresponding layers, thus reducing the energy of superlattices. An increase in the density of Jahn–Teller Mn$^{3+}$ ions in the L0 layers results in the deepening of potential wells inside these layers, bringing their states closer to the R1 states. This suggests an intense exchange of electrons between the L0 and R1 layers at $T$ = 22 K, which makes electron densities in these layers equal (Fig. 2b). A new state of superlattices arises, which is long-lived at low temperatures, since it does not disappear for at least 20 min after the end of pumping.

In Fig. 2c, we demonstrate a sequence of FMR lines for the same sample after 16 h of natural heating up to 250 K (after optical pumping) and its subsequent cooling down to 17.5 K (without new optical pumping). Black and red points and lines correspond to increasing and decreasing magnetic fields, respectively. We can see that the intensities of the L2 and L1 lines still do not change at increasing and decreasing fields. At the same time, as the field increases, the intensities of the R1 and R2 lines increase steeply and exceed the intensity of the L0 line. Thus, the state forming after optical pumping is not recovered when the sample is cooled down to 17.5 K after its heating and subsequent cooling. This means that the state arising after optical pumping with the same populations of the L0 and R1 layers (Fig. 2b) is long-lived.

It is natural to assume that the heating of the sample after pumping leads to a certain redistribution in the L0, R1, and R2 layers of excess electrons generated by optical pumping, which leads to the reduction of the superlattice energy. The heating of the sample should result in an increase in the kinetic energy of electrons, thus causing their redistribution among the layers owing not only to tunneling between the L0 and R1 layers (Fig. 2b) but also to the hopping conductivity, which favors overcoming higher barriers at the boundaries of other layers of superlattices. The population of the deeper potential wells with the maximum number of $Mn^{3+}$ ions is the most probable. This is the case for the R2 layer, which is nearly empty at low temperatures (Figs. 2a, 2b). At the same time, two other layers, R1 and L0, are populated but with a lower probability. Structural distortions also increase in them owing to an increase in the number of Jahn–Teller $Mn^{3+}$ ions, for which the potential wells in these layers are deeper. In this case, all excess electrons are localized at $Mn^{3+}$ ions. As a result, intense narrow FMR lines related to the R2, R1, and L0 layers are observed (Fig. 2c, black points and curves). With a subsequent decrease in the magnetic field, the intensities of the L0, R1, and R2 lines decrease steeply (Fig. 2c, red curves). In this case, the R2 line intensity becomes nearly equal to its value in dynamical equilibrium before pumping, whereas the R1 line intensity exceeds that of the L0 line, although only slightly. Only an additional increase in the magnetic field recovers a state similar to that corresponding to the initial equilibrium state of the superlattice before pumping (Fig. 2d), but with slightly lower intensities of the FMR lines. The cycling of the magnetic field, recovering the ferromagnetic orientation of the layers in superlattices, enhances the double exchange, leading to the tunneling of electrons between the layers in superlattices. It turns out that the resulting state exists up to a higher temperature than the equilibrium state before optical pumping (Fig. 3).

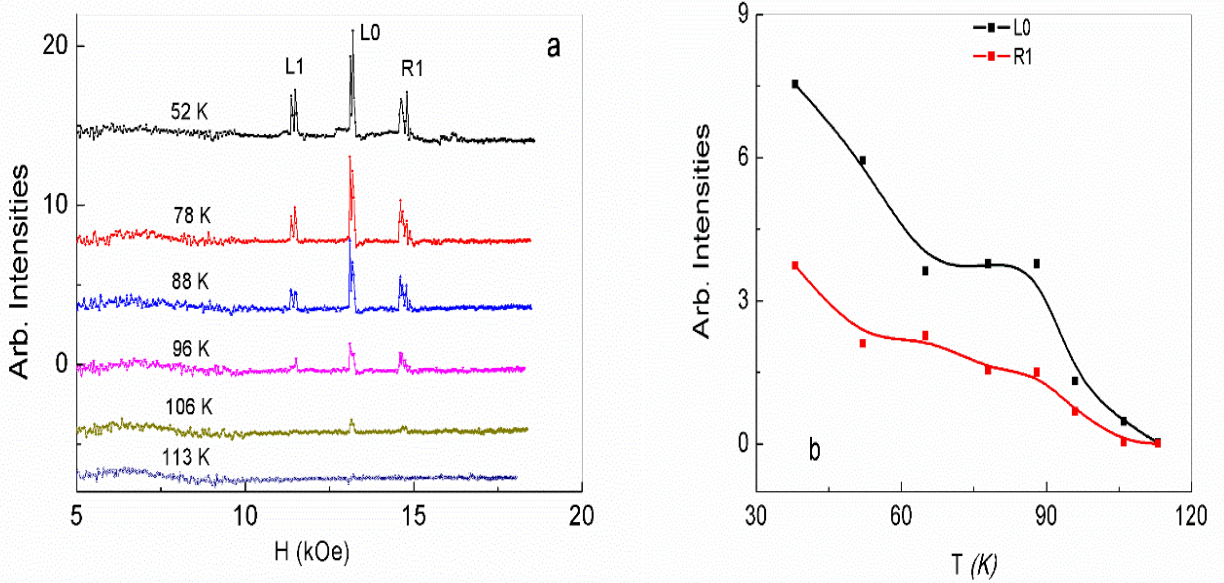

**Fig. 3.** (a) Sequence of FMR lines corresponding to the layers in superlattices after optical pumping and cycling in the applied magnetic field at various temperatures. Actually coinciding zeros of intensity are displaced relative to each other.
(b) Temperature dependence of intensities for the L0 and R1 lines.

In Fig. 3, we show the set of the FMR lines in the recovered equilibrium state after optical pumping at various temperatures (Fig. 3a) and the temperature dependence of the intensities of the L0 and R1 lines in this state (Fig. 3b). We can see that the equilibrium states of superlattices after optical pumping exist up to $T \approx 115$ K, whereas before pumping, they are observed up to $T \approx 60$ K [12].

As mentioned above, more intense and narrower L1, L0, and R1 lines in Fig. 2 are doubled: the low intensity wide satellites are observed near these lines at somehow lower fields. This occurs because the initial crystal contains not only $Ce^{4+}$ ions but also $Ce^{3+}$ ions, although in a smaller number, which induce local regions with strong lattice distortions because of the presence of alone $6s^2$ electron pairs in their outer shells. We previously observed this effect in $R_{0.8}Ce_{0.2}Mn_2O_5$ (R = Er and Tb) [27]. In these regions with a certain number of localized electrons, superlattices with slightly different parameters are formed.

To summarize, it has been demonstrated for the first time that the states of 1D superlattices (ferromagnetic semiconductor heterostructures) in $Eu_{0.8}Ce_{0.2}Mn_2O_5$ multiferroics, which at sufficiently low temperatures are phase separation domains, can be controlled by optical pumping and applied magnetic field. Optical pumping leads to the formation of a new long-lived state of superlattices with a different distribution of $Mn^{3+}$ and $Mn^{4+}$ ions and of electrons changing their charge in these layers. The cycling of the applied magnetic field in this new state of superlattices after pumping recovers the state with the distribution of intensities of resonance lines similar to that existing before pumping. In this case, the intensities in the sequence of resonance lines recorded after pumping are slightly lower than those before pumping. This is due to the increased densities of electrons and $Mn^{3+}$ and $Mn^{4+}$ ion pairs in the layers. However, these lines exist up to a higher temperature, which means that the new equilibrium state recovered by the magnetic field after optical pumping is more favorable in energy than the state before pumping.

**FUNDING**

This work was supported by the Russian Foundation for Basic Research (project no. 18-32-00241) and by the Presidium of the Russian Academy of Sciences (program 1.4 "Topical Problems of Low-Temperature Physics").